\documentclass[aps,prb,twocolumn,nobibnotes]{revtex4-1}
\usepackage{graphics,graphicx,amsfonts,amsmath,amsbsy,amssymb,color}
\usepackage{bm}
\usepackage[a4paper,vmargin={20mm,20mm},hmargin={20mm,10mm}]{geometry}
\usepackage{subfigure}
\usepackage{paralist}
\usepackage{bbold}
\usepackage[inline]{enumitem}

\def \beq {\begin{eqnarray}}
\def \eeq {\end{eqnarray}}

\newcommand{\ket}{\rangle}

\begin{document}
\title{Rigorous wave function embedding with dynamical fluctuations}
\author{Edoardo~Fertitta}
\author{George~H.~Booth}
\email{george.booth@kcl.ac.uk}
\affiliation{Department of Physics, King's College London, Strand, London, WC2R 2LS, U.K.}

\begin{abstract}
The dynamical fluctuations in approaches such as dynamical mean-field theory (DMFT) allow for the self-consistent optimization of a local fragment, hybridized with a true correlated environment. We show that these correlated environmental fluctuations can instead be efficiently captured in a wave function perspective in a computationally cheap, frequency-independent, zero-temperature approach. This allows for a systematically improvable, short-time wave function analogue to DMFT, which entails a number of computational and numerical benefits. We demonstrate this approach to solve the correlated dynamics of the paradigmatic Bethe lattice Hubbard model, as well as detailing cluster extensions in the one-dimensional Hubbard chain where we clearly show the benefits of this rapidly convergent description of correlated environmental fluctuations.
\end{abstract}
\date{\today}
\maketitle

\section{Introduction}

Reliable computational probes of correlated quantum systems are key to our understanding of competing phases of matter and progress in materials science. Developments in recent years have dramatically improved the tools at our disposal, however, their ability has long been stymied by the difficulty in reaching the thermodynamic 'bulk' limit of system size. It is in this regard that recent `quantum cluster' paradigms have come to the fore\cite{cluster1,cluster2,cluster3,PhysRevLett.109.186404}. These approaches rely on a mapping between the bulk system and a simplified quantum model which represents a fragment of the system (or `impurity'), and its coupling to its surrounding environment. This simplified locally correlated quantum problem can be solved to high accuracy with a number of techniques before a self-consistency is used to update the original system\cite{Gull07,PhysRevB.90.085102,PhysRevB.96.085139,PhysRevB.95.045103}. These approaches have recently become key tools to describe lattice models\cite{PhysRevB.97.075112,Fan1951,Zheng2016}, materials science\cite{Park2014,Leonov2015}, and even quantum chemistry\cite{Wouters2016,Weber5790,Chen2015,PhysRevLett.106.096402}. 
In this work, we present a step forwards in quantum cluster approaches, with the development of a technique which combines the advantages of a number of existing methods, and interpolates between the physics they capture. We demonstrate the ability to encompass true dynamical correlation effects in challenging correlated lattice models, whilst crucially maintaining a cheap, static mapping to a finite quantum impurity problem.

The most widespread quantum cluster framework is given by dynamical mean-field theory (DMFT), where the coupling of the impurity to the rest of the system is described by temporal quantum fluctuations\cite{cluster3}. This framework is centred on a Green function description, with a local self-energy describing the correlated physics of the impurity. More recently, an alternative wave function approach was developed called density matrix embedding theory (DMET). In this, the coupling to the environment is described by the entanglement between the impurity and the rest of the system, which can be expressed exactly by the span of a set of physical degrees of freedom\cite{PhysRevLett.109.186404,doi:10.1002/9781119129271.ch8}.
While clarifying differences in the physics of these two approaches, we will develop an improved quantum cluster method which combines many of the strengths of both DMFT and DMET, and results in a computationally efficient, static method, which can be systematically improved to include all of the physics arising from the self-consistent temporal fluctuations considered within the DMFT framework, and which are beyond the standard DMET approach.

Before describing this new approach, it is helpful to further consider the properties of DMFT and DMET from which it is inspired.
In DMFT, the central object is the local Greens function of the impurity space, which is coupled to its environment via a hybridization. This is consequently modelled via a (formally infinite) Anderson impurity model. The self-consistency takes place via a local self-energy over the impurity, which is updated to match the correlated local Green function of the cluster to the one from the lattice. 
By analogy, in DMET the central object is the one-body reduced density matrix (RDM), $\langle {\hat c}^{\dagger}_{\alpha} {\hat c}_{\beta} \rangle$, where $\{\alpha, \beta\}$ denote the $n_{\rm imp}$ impurity sites. This object is also the static limit Green function as $\tau \rightarrow 0^+$. 
By analogy to the local dynamic self-energy in DMFT, DMET then attempts to match these via a local one-body static potential, which can be considered a high-frequency static limit of a self-energy.
Because of this, DMET has sometimes been (erroneously) considered as a `static' limit of DMFT, but this is not accurate and stems from a fundamental limitation of DMET. 

The DMET self-consistency involves finding a one-particle (hermitian) potential spanning the impurity space, denoted the `correlation potential'. This is designed to match the impurity part of the lattice RDM to the correlated RDM resulting from the solution of the quantum cluster model. However, this is (in general) not possible, since a correlated RDM is not non-interacting $v$-representable, as pointed out by a number of researchers before\cite{Tsuchimochi2015,PhysRevB.96.235139}. The self-consistency is therefore written as a minimization of the difference between the subspace lattice RDM and the correlated impurity RDM in a least-squares sense\cite{Wouters2016}. 

This correlation potential imposes that the lattice model is always represented as a single Slater determinant, which while important for the efficiency of the method, ensures that no `true' correlated (e.g. Mott) physics is returned to the full lattice solution through the self-consistency (as is done through the dynamical nature of the self-energy in DMFT)\cite{Gunst1951,PhysRevB.91.155107}. The optimized lattice solution can therefore only attempt to mimic the correlated physics of the impurity cluster via static symmetry-breaking, with all non-trivial dynamical character of the cluster solution lost on return to the lattice. Physics beyond symmetry-breaking is therefore inaccessible, whilst in the single-site approximation no change in the lattice description is possible at all. Furthermore, infinite-dimensional lattices are no longer exact within DMET (whilst exact in DMFT), as their correlated phase transitions are driven entirely by the dynamical nature of the local physics\cite{cluster3}.

While these limitations mean that only in a loose sense can DMET be described as a static approximation to DMFT\cite{PhysRevB.96.235139}, it does not mean that the results are necessarily of worse quality. Indeed, DMET possesses a number of computational advantages which have enabled it to reach or surpass the accuracy of DMFT, achieving some of the most accurate results available for the Hubbard model\cite{PhysRevX.5.041041}, and application in quantum chemistry\cite{Wouters2016}. This is due to the relatively simple ground-state cluster problem which results, and the highly efficient mapping from the lattice. 
As the lattice state is constrained to be written as a single Slater determinant, $|\phi\rangle$, the entanglement between the impurity and its environment can be exactly exposed via a Schmidt decomposition as single-particle `bath' degrees of freedom whose dimensionality is the same as that of the impurity space\cite{doi:10.1002/9781119129271.ch8}. This results in a unique projection from the lattice to a finite, compact quantum cluster model which allows for a relatively efficient ground-state solution compared to the requirement in DMFT for the full correlated impurity Green function with a retarded coupling or formally infinite bath. 

This has enabled accurate and efficient, zero-temperature, ground-state impurity solvers to be used, which has allowed large impurity cells with high momentum resolution to be considered\cite{Zheng2016,PhysRevB.95.045103,Zheng1155}. This has even allowed for long-range inhomogeneous phases such as stripes to be found, which are traditionally hard in cluster models\cite{Zheng1155}. 
Furthermore, the wave function description has also allowed for traditionally inaccessible quantities to be easily extracted. These large advantages have meant that despite its limitations, it has nevertheless proved an important tool for strongly correlated extended systems.

The approach proposed here extends the DMET methodology to allow for a systematic improvement in the resolution of the {\em effective} treatment of energy-dependent (temporal) fluctuations in the lattice. This results in a true correlated lattice through the self-consistency, beyond an effective single determinant treatment. This is achieved whilst remaining in an entirely static and inexpensive formulation, requiring only the ground-state solution of the resulting algebraically-constructed finite cluster model at each step. This leads to a method which can interpolate between the physics of DMET and DMFT, without sacrificing the efficient aspects of DMET which has enabled its success. We demonstrate that even in the single-site approximation, we can now describe correlation-driven quantum phase transitions in the lattice without the need for symmetry-breaking, and the ability to systematically converge to exact infinite-dimensional results, as well as a treatment of multiple impurity clusters.

\section{Method}

In order to surpass the limitation of DMET and introduce a systematically improvable description of energy-dependent fluctuations, the proposed approach relies on the self-consistent matching of not just the single-particle RDM, but the impurity hole and particle $n$th-order {\em energy-weighted} density matrices, \mbox{${\tilde {\bf T}_h^{(n)}}$} and \mbox{${\tilde {\bf T}_p^{(n)}}$} respectively, defined from the lattice solution as
\begin{eqnarray}
\tilde {\bf T}_h^{(n)} &=& P_{\rm imp} {\bf C_{\rm hole}} (h - e_0)^n {\bf C_{\rm hole}^{\dagger}} P_{\rm imp}	\label{eqn:latmom_h} \\	
\tilde {\bf T}_p^{(n)} &=& P_{\rm imp} {\bf C_{\rm part}} (h - e_0)^n {\bf C_{\rm part}^{\dagger}} P_{\rm imp}	\label{eqn:latmom_p}
\end{eqnarray}
where $h$ is a single-particle lattice Hamiltonian with ground state energy \mbox{$e_0$},
${\bf C_{\rm hole/part}}$ represent the occupied (hole) and unoccupied (particle) eigenstates of $h$, and $P_{\rm imp}$ is the projector onto the chosen impurity sites.
These mean-field quantities can be matched to their counterparts from the correlated solution of the quantum cluster model which can be computed  as expectation values of the correlated ground state many-body wavefunction $\Psi_0$
\begin{eqnarray}
T_{h, \alpha \beta}^{(n)}=\langle \Psi_0 | c_{\alpha}^{\dagger} (\hat H_{\rm clust} - E_0)^n c_{\beta} | \Psi_0 \rangle\\
T_{p, \alpha \beta}^{(n)} = \langle \Psi_0 | c_{\alpha} (\hat H_{\rm clust} - E_0)^n c_{\beta}^{\dagger} | \Psi_0 \rangle,
\end{eqnarray}
where $E_0$ is the correlated ground state energy of the cluster model with Hamiltonian ${\hat H_{\rm clust}}$.

Despite being purely static quantities, these expectation values can be rigorously related to the character of the {\em dynamical} single-particle spectral function of the system, as they define the $n$-th order moments of the individual hole and particle distributions of the spectral functions, with 
\begin{eqnarray}
T_{h, \alpha \beta}^{(n)} = -\frac{1}{\pi}\int_{-\infty}^{\mu} \Im[G_{\alpha \beta}(\omega+i0^+)] \omega^n d\omega  \\
T_{p, \alpha \beta}^{(n)} = -\frac{1}{\pi}\int_{\mu}^{\infty} \Im[G_{\alpha \beta}(\omega+i0^+)] \omega^n d\omega ,
\end{eqnarray}
where ${\bf G}(\omega)$ is the Green function, and $\mu$ is the chemical potential. The zeroth (hole) moment is simply the RDM (used for the DMET self-consistency) defining the integrated weight of the hole distribution, while the higher moments build in an increasingly well-resolved dynamical character by defining the mean, variance, skew, bimodal and beyond character of these individual particle and hole distributions.

On the other hand, in imaginary-time they can also be related to the Taylor expansion of the Green function for short hole/particle propagation times, as 
\begin{eqnarray}
{\bf T}_{h}^{(n)} &=& \left. \frac{d^n {\bf G(\tau)}}{d \tau^n}\right|_{\tau=0^-} \\
{\bf T}_{p}^{(n)} &=& (-1)^{(n+1)}\left. \frac{d^n {\bf G(\tau)}}{d \tau^n}\right|_{\tau=0^+}.
\end{eqnarray}
These therefore describe the weights of the paths of increasing length for holes or electrons which leave and return to the impurity space.
This expansion is well-defined for all $n$, and converges in the \mbox{large-$n$} limit to a unique description of the full dynamics of the single-particle propagator.
It should be pointed out that, while this is seen as a short-time expansion of the Green functions, it is not a high-energy expansion. This is because (in contrast to the usual definition of the spectral moments) the particle and hole moments are separately considered, rather than their sum in the full central moment\cite{Potthoff98,Potthoff01}. This allows true low-energy (e.g. Mott) physics to be described on the lattice at lower orders than constraining the central moments of the spectrum.

Although a self-consistent matching of energy-weighted density matrices between the lattice and cluster model will allow for dynamical correlated effects to be returned from the cluster to the lattice, a local static `correlation' potential as optimized within DMET ($v_c$) is insufficient to rigorously match even the RDM (0th hole moment), and this becomes increasingly true for higher moments\cite{Tsuchimochi2015,PhysRevB.96.235139}.
To overcome this, we supplement $v_c$ with a set of $n_{\rm aux}$ auxiliary single-particle degrees of freedom, which couple to the impurity space on the physical lattice, and to all symmetry-equivalent $n_{\rm cells}$ impurity cells. 
The auxiliary hamiltonian can be written as
\begin{equation}
h_{\rm aux}=\sum_{x=0}^{n_{\rm cells}-1} \sum_{\alpha}^{n_{\rm imp}} \sum_{k}^{n_{\rm aux}} \nu_{\alpha k} ({\hat c}^{\dagger}_{\alpha+x} {\hat c}_{k+x} + h.c.) + \varepsilon_{k} {\hat c}^{\dagger}_{k+x} {\hat c}_{k+x}	,
\end{equation}
where ${\hat c}_{\alpha+x}$ denotes the annihilation operator on the site $\alpha$ after translation by $x$ impurity cells. These auxiliary sites have energies ($\varepsilon_{k}$) and couplings ($\nu_{\alpha k}$) to the physical impurity cell which are fit to optimally mimic the correlation-driven dynamical fluctuations in the lattice arising through the self-consistency scheme. This expands the single-particle lattice hamiltonian, \mbox{$h\rightarrow h + h_{\rm aux}$}.  Increasing $n_{\rm aux}$ only increases the computational cost of the lattice diagonalizations and does not affect cluster size or effort of the solver for the resultant projected `cluster' problem.

\begin{figure}[t]
\includegraphics[scale=0.5,trim={0 0 0 0},clip]{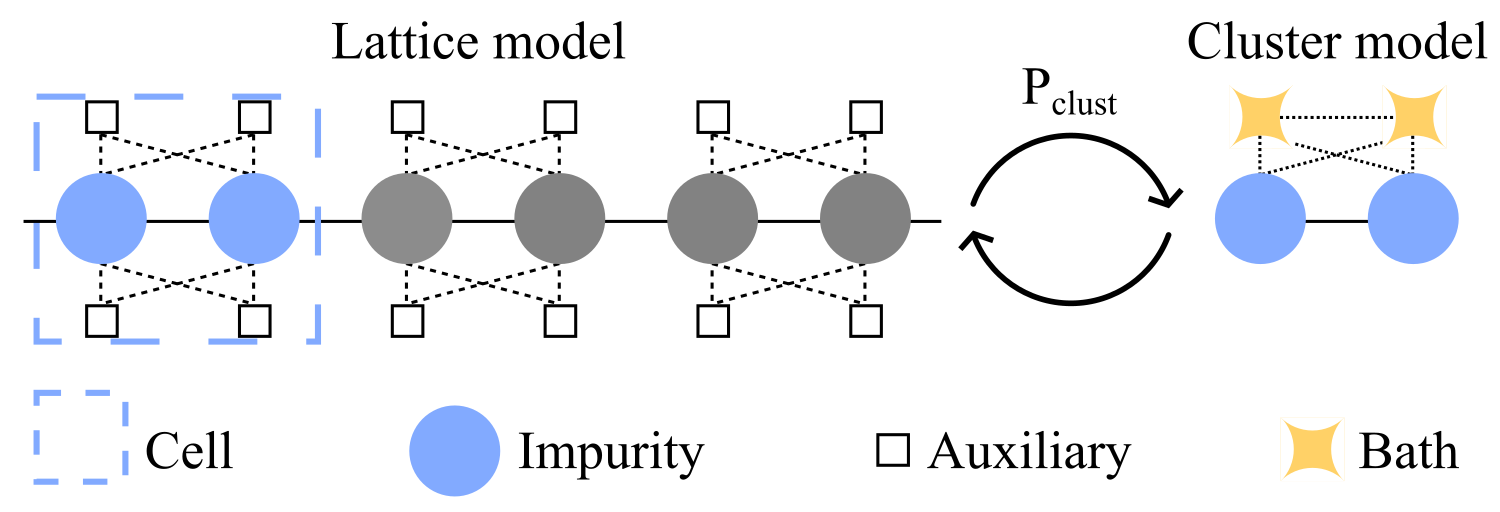}
\caption{Overview of the embedding scheme for a 1D lattice expanded with auxiliary orbitals. The exact projection into the cluster space $P_{\rm clust}$, results in a static, finite quantum problem which contains the chosen two-site impurity cell, and the bath space required to span the exact lattice dynamical fluctuations in the particle and hole sectors up to a desired order.}
\label{fig:Ham_scheme}
\end{figure}

The use of an enlarged single-particle space to describe dynamical character due to a self-energy has been used before, including non-equilibrium DMFT and exact-diagonalization solvers where the hybridization is expressed via a non-interacting bath space\cite{PhysRevB.89.035148,Liebsch2012}. In this way, the effect of \emph{any} dynamical self-energy (required to match the lattice and cluster moments) can be expressed by increasing the number of static auxiliary orbitals which couple to the physical lattice. This correlated lattice, and the projection between the lattice and cluster model for a two-site impurity cell is shown schematically in Fig.~\ref{fig:Ham_scheme}.

In a similar fashion to DMET, the construction of \mbox{$\hat H_{\rm clust}$} is carried out by projecting the lattice Hamiltonian into the cluster space $P_{\rm clust}$ which is analytically constructed by means of the Schmidt decomposition of the enlarged lattice $h$.
In this form, the entanglement between the spaces given by the original wave function is exposed as a linear fluctuation space which has the maximum number of dimensions as the smaller of the two spaces. As discussed extensively in Refs.~\onlinecite{Wouters2016,doi:10.1002/9781119129271.ch8,PhysRevB.89.035140}, significant simplifications emerge if the state to be decomposed can be written as a single-particle state, given by the solution to a quadratic Hamiltonian, or the action of any single-particle operator on that solution.
However, in contrast to DMET where the bath orbitals are constructed by Schmidt decomposing a single Slater determinant, in order to ensure that the cluster spans the full fluctuation space required to represent {\em all} desired ${\tilde {\bf T}_{h/p}^{(n)}}$ from the lattice faithfully within the cluster model, it is necessary to augment the bath space by decomposing wave functions of the form
\begin{equation}
h^m c_{\alpha}^{(\dagger)} |\phi \rangle	,
\end{equation}
where $|\phi \rangle$ represents a single Slater determinant spanning the lattice (and auxiliary) sites, and $\alpha$ represents an impurity site. The fluctuation space into the rest of the lattice from these states define the bath orbitals which are required to ensure that the particle and hole moments are matched under the action of $h$ by construction in both the lattice and the cluster. Performing the Schmidt decomposition gives a single degree of freedom for each $\alpha$, expressed in the lattice space orthogonal to the impurities as
\begin{equation}
|b_{\alpha,m}^{\rm hole} \rangle = \frac{1}{{\sqrt{N_{\alpha, m}^{\rm hole}}}} \sum_{t \notin {\rm imp}} \left( \sum_{\epsilon_i<\mu} \epsilon_i^m C_{\alpha i} C_{t i}^* \right) {\hat c}_t^{\dagger} ,
\end{equation}
for the bath orbitals of the hole moment, and
\begin{equation}
|b_{\alpha,m}^{\rm particle} \rangle = \frac{1}{{\sqrt {N_{\alpha, m}^{\rm particle}}}} \sum_{t \notin {\rm imp}}  \left( \sum_{\epsilon_i>\mu} \epsilon_i^m C_{\alpha i} C_{t i}^* \right) {\hat c}_t^{\dagger} ,
\end{equation}
for the bath orbitals of the particle moment, where $t$ denotes degrees of freedom in the lattice orthogonal to the impurity space, $N$ values represent the normalization constants of the resulting orbitals, and the $\{ {\bf \epsilon} ; {\bf C} \}$ pairs represent the eigenvalues and vectors from the diagonalization of the lattice hamiltonian ($h$) of the system, with $\mu$ as the chemical potential of this hamiltonian.

It should be noted that the bath orbitals for different $\alpha$ and $m$ are not necessarily orthogonal, and so an orthogonalization scheme and removal of linear dependencies which necessarily arise is necessary. In order to ensure that the cluster model spans the hole and particle moments up to order $n$, it is necessary to span all wave functions of the type \mbox{$h^{ \lceil \frac{n-1}{2} \rceil} c^{(\dagger)}_{\alpha} | \phi \ket$}. This reduction from the full set of $n$ wave functions is due to a Wigner `$2n+1$' rule which ensures that we only need the wave functions (and hence bath orbitals) correct up to order $\frac{n-1}{2}$ to ensure the full lattice moments can be faithfully represented in the cluster up to order $n$\cite{Wigner80}. 
 
The result is that the number of bath orbitals now grows linearly with both the size of the impurity space (as in DMET which this reduces to in the $n=0$ limit) and the order through which the lattice dynamical effects are correct. This gives a strict upper bound of $n_{\rm imp}(2\left \lceil \frac{n-1}{2} \right \rceil + 1)$ bath orbitals in the cluster. However, due to further `accidental' linear dependencies in the bath orbitals, in practice, the number of bath orbitals is frequently less than this upper bound. The bath orbitals also can be seen to contribute either two or zero electrons to the cluster depending on whether they are derived from the particle or hole character state. One immediate consequence of this decomposition is that the bath space required to represent $n=1$ is identical to $n=0$, and therefore {\em no} additional computational effort is required in the solver to add the first moment dynamical character in the lattice compared to normal DMET where only the $n=0$ dynamical effects are captured.

We describe the final algorithm for an example Hubbard lattice model consisting of hopping (${\hat H}_t$) and interaction (${\hat H}_U$) terms, as
\begin{eqnarray}
H &=& {\hat H}_t + {\hat H}_U	,	\\
&=& -t \sum_{\langle i,j \rangle, \sigma} ({\hat c_{i,\sigma}^{\dagger}} {\hat c_{j,\sigma}} + {\hat c_{j,\sigma}^{\dagger}}{\hat c_{i,\sigma}}) + U\sum_{i=1}^N {\hat n_{i \uparrow}}{\hat n_{i \downarrow}}
\end{eqnarray}
where $\langle i,j \rangle$ represents summation over nearest-neighbor lattice sites and ${\hat n_{i \sigma}}={\hat c_{i \sigma}^{\dagger}}{\hat c_{i \sigma}}$ is the spin-density operator for spin $\sigma$ and site $i$.
Once an impurity space of $n_{\rm imp}$ sites with corresponding impurity projector $P_{\rm imp}$ is selected, and maximum moment order of $n$ is chosen, the final algorithm is as follows:
\begin{enumerate}[label=(\roman*)]
\item The bath orbitals required to represent the ${\tilde {\bf T}_{h/p}^{(n)}}$ hole/particle moments from a single-particle lattice hamiltonian, $h$ (initially taken to be ${\hat H}_t$), are constructed as shown above. The set of these orbitals augment the impurity degrees of freedom to form a projector into the cluster space of impurity + bath orbitals, $P_{\rm clust}$.
\item A chemical potential ($\mu_{\rm bath}$) is used to ensure the correct number of electrons in the impurity space of the cluster, while the total number of electrons in the cluster is given by the projection of $|\phi \rangle$ into the cluster, and will be an integer by construction.
\item The interacting cluster Hamiltonian is formed as
\begin{eqnarray*}
\hat H_{\rm clust} &=& P_{\rm clust} h P_{\rm clust}\\
                   &+& P_{\rm imp}({\hat H}_U - v_c - h_{\rm aux} - \mu_{\rm bath} I) P_{\rm imp},
\end{eqnarray*}
where $h_{\rm aux}$ and $v_c$ are initially set to zero. This form ensures that the auxiliary sites representing the effective local self-energy are removed from the correlated impurity space.
\item An accurate solver (exact diagonalization in this work\cite{pyscf}) is used to compute $\Psi_0$, $E_0$ and the correlated expectation values ${\bf T}_{h/p}^{(n)}$.
\item The parameters defining $v_c$ and $h_{\rm aux}$ (energies $\varepsilon$ and couplings $\nu$ for the $n_{\rm aux}$ auxiliary sites) are optimized by minimizing the squared moment error between the lattice and cluster moments as
\begin{equation*}
C = \sum_{\alpha \beta}^{n_{\rm imp}} \sum_{z=p/h} \sum_{n'=0}^n w_{n'} ({\tilde T}_{z, \alpha \beta}^{(n')}[v_c,\varepsilon ,\nu ]-T_{z, \alpha \beta}^{(n')})^2	,
\end{equation*}
where $w_{n'}$ are weighting parameters of the moments, to account for their growing absolute value, which we take to be $w_{n'}=\frac{1}{n'!}$
\item The updated $v_c$ and $h_{\rm aux}$ are replicated for all translationally equivalent impurity spaces in the lattice, in the same fashion as the self-energy in DMFT or $v_c$ in DMET. This enlarged lattice is then used as the new lattice hamiltonian $h$, and we return to step (ii) until convergence.
\end{enumerate}

The numerical minimization of the cost function $C$ with respect to all mean-field physical and auxiliary degrees of freedom is performed using analytic gradients with respect to all adjustable variables (given by $v_c, \nu_{\alpha k}$ and $\epsilon_k$). Additional consideration needs to be paid to the choice of the global number of electrons in the modified lattice augmented with the auxiliary sites, as this also can change through the optimization. This discrete optimization over electron number is changed to a continuous optimization by working in a grand-canonical ensemble at a very low but finite temperature for the computational of the moments of the hole/particle states on this lattice for this step. This small fictitious temperature mean that powers of the Fermi-Dirac distribution are used in the calculation of the lattice moments given by Eq.~\ref{eqn:latmom_h} and \ref{eqn:latmom_p}, and the sum extended to all lattice eigenstates. This changes the optimization of the total electron number to the optimization of an additional continuous variable given by the chemical potential of these distributions. The lattice moments are then reevaluated at zero temperature in order to form the bath orbitals. 

This approach also avoids any difficulties due to degeneracies at the Fermi level of the lattice, which give rise to divergent gradients in the optimization. This approach to a small temperature in fitting is shared with (formally zero-temperature) approaches in exact-diagonalization DMFT, where despite solving the zero-temperature correlated problem, the fitting of bath orbitals on the Matsubara axis necessarily introduces temperature effects and smears occupation about the Fermi level. In this work, this fitting temperature is set to $\beta=150t$, and has negligible effects of the solution found at this low temperature, but ensures a fit of only continuous variables.
Finally, the systems considered in this work are all at particle-hole symmetric points, which we exploit to reduce the number of independent parameters required to fit the auxiliary space by a factor of two.

The method as described above (which we denote {\em Energy-weighted Density Matrix Embedding Theory}, EwDMET) allows for coarse-grained dynamical effects in the lattice to be self-consistently optimized, in a systematically-improvable fashion. In DMET, the explicit correlated dynamical character of the impurity is lost on return to the lattice picture, while in DMFT it is returned via a self-energy. In this approach, we can converge to the capture the entire dynamics (and hence physics of DMFT) by increasing $n$ and $n_{\rm aux}$, which can be considered as controlling the resolution of the dynamical character of $G(\omega)$ and $\Sigma(\omega)$ respectively in the DMFT picture, all whilst retaining a fully static wave function framework. The limit of $n = n_{\rm aux} =0$ returns DMET exactly, while $n \rightarrow \infty, n_{\rm aux}\rightarrow\infty$ returns the zero-temperature physics of DMFT, and we can interpolate between these two limits as desired. 

However, this approach also interpolates between the physics of another quantum cluster approach; the `two-site DMFT' proposed by Potthoff\cite{Potthoff01}. In this, a single bath orbital is used to capture the self-consistent hybridization of a single impurity by describing the zeroth-order hole distribution, and the second {\em central} moment of the Green function. This can be shown to be equivalent to optimizing up to the first hole and particle moments separately, which is equivalent to that achieved (in a `static' context) for the EwDMET approach with $n_{\rm imp}=1$ and $n=1$. This would then automatically give an explicit single bath orbital, with a cluster problem of the same size as in the approach of Potthoff in Ref.~\onlinecite{Potthoff01}. This would capture the same local physics of the system, while cast in an explicit wave function formulation. In contrast, the EwDMET method allows for systematic improvability by increasing both the number of moments, and impurity cluster size, whilst retaining the computationally efficient wave function formalism. As a final note, it is worth stressing that the $n_{\rm imp}\rightarrow \infty$ will return exact results, in common with both DMET and DMFT. 

We now turn to numerical investigation of the benefits of including this dynamical lattice character, with applications to the infinite- and 1-dimensional limits of the Hubbard model.

\section{Infinite-dimension Bethe Hubbard Lattice} 

\begin{figure}[b!]
\includegraphics[scale=0.49,trim={5 0 0 0},clip]{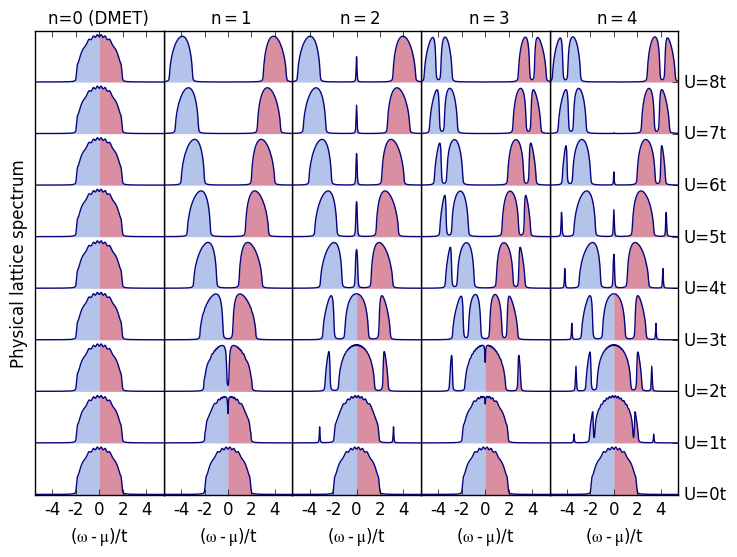}
\caption{Single-particle spectral function of the physical lattice for the half-filled Hubbard Bethe model with increasing hole/particle moment self-consistency. The number of auxiliary sites which augment the (physical) lattice hamiltonian is given by $n_{\rm aux}=n$, which is sufficient to exactly match the hole/particle moments (denoted by the blue/red color respectively). Metal-to-Mott insulator behaviour as well as emergent Kondo physics is seen in the lattice with increasing dynamical resolution as the $n$ increases, in keeping with DMFT results. DMET ($n=0$) is unable to change the character of the lattice spectral function through the self-consistency.}
\label{fig:BetheLattice}
\end{figure}

\begin{figure}[h!]
\includegraphics[scale=0.5,trim={0 0 0 0},clip]{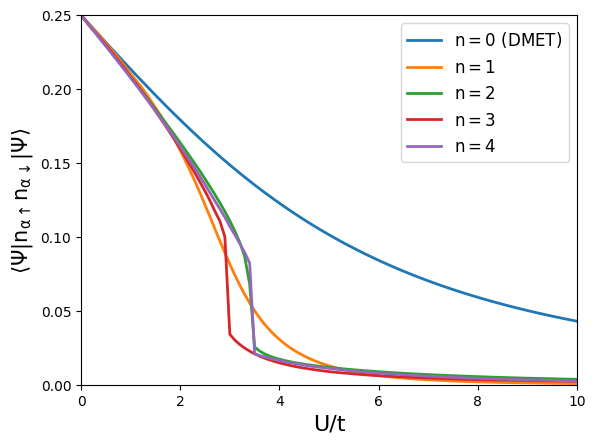}
\caption{Double occupancy of the impurity site from the correlated cluster solution. Results are shown for different self-consistent moment order $n$ for the Hubbard Bethe lattice, with a MIT observed at the discontinuity for $n \geq 2$.}
\label{fig:Bethe_DO}
\end{figure}

The infinite-dimensional Hubbard Bethe lattice has been a cornerstone in the development of DMFT. This is because it was shown that despite a highly non-trivial phase diagram including a finite $U$ metal to Mott insulator transition (MIT), single-impurity DMFT was exact due to the local (yet highly dynamic) nature of the self-energy\cite{PhysRevLett.62.324,PhysRevLett.70.1666}. This system is still extensively used and debated\cite{PhysRevB.73.205121}, and therefore represents an ideal test as to whether this correlated behaviour can be captured in this formally static approach, both in the cluster and lattice descriptions of the system. These two descriptions will not necessarily be identical at convergence, as the only constraint is that their separate particle and hole moments up to a given order will be identical, assuming that $n_{\rm aux}$ is sufficient to describe them. The non-interacting lattice is defined by a semi-circular density of states, which is fit to an initial lattice Hamiltonian, consisting of 100 orbitals in a star geometry surrounding an single central (impurity) site\cite{Kollar02}. A Hubbard $U$ term is then included on all sites to define the interacting Hamiltonian in which this impurity is self-consistently embedded. The EwDMET procedure above was run for different values of $n$ and $n_{\rm aux}$, and it was found that $n_{\rm aux}=n$ was sufficient to exactly fit all $n$ moments, with further auxiliary orbitals redundant.

Fig.~\ref{fig:BetheLattice} shows the lattice density of states (DoS), and demonstrates the improved lattice description as the hole/particle moment self-consistency is increased, thereby more finely resolving the effective correlation-driven dynamical character. Self-consistency just on the zeroth moments is equivalent to DMET (leftmost plots), and simply constrains the integrated weight of the distributions to be correct (number of electrons/holes). However, the lattice also shows no change with $U$, giving the non-interacting semi-circular DoS at all values. This is to be expected, since the zeroth hole moment defined the occupancy of the correlated impurity, which is constrained to be half-filled. No further information is returned from the correlated solution of the impurity problem to the lattice in this case, and so no correlation effects can be found. Note that constraining this zeroth-order hole moment automatically fulfils the zeroth-order particle moment, since these must sum to unity, and therefore no further information is contained in the zeroth-particle moment (this is not true for higher $n$ values).

 The $n=1$ results constrain the mean of the hole/particle distributions (while requiring only the same single bath orbital as $n=0$ DMET in the cluster solution). This added self-consistency ensures that a Mott gap opens in the lattice. This is achieved by adding an auxiliary site per impurity at the Fermi level to suppress spectral weight on the physical lattice at this energy. $n=2$ ensures additional matching of the variances of each of these distributions, and leaves a sharp Kondo-like peak at the Fermi level. However, this only disappears at infinite $U$. $n=3$ introduces structure in the Hubbard bands by enforcing the skew of each distribution. However, the full qualitative character of the exact solution is found at $n=4$, with structured Hubbard bands and a sharp Kondo peak which disappears at finite $U$ (around $U/t=6$). This value compares well with real-frequency DMFT and QMC results\cite{PhysRevLett.70.1666,Gros94,PhysRevB.90.085102} for this correlation-driven metal-to-insulator transition. Increasing the moments further to $n=5$ and $6$ yields no substantial changes from the qualitatively correct $n=4$ depiction of the lattice density of states, indicating that the broad trends of the model are converged at this point.

We can also consider the solution in the cluster, where the transition point can be seen from the double occupancy on the impurity site (Fig.~\ref{fig:Bethe_DO}). Here, the DMET solution ($n=0$) is quite incorrect in the large-$U$ limit, which is improved by the $n=1$ constraints. However, neither of these solutions exhibit a discontinuity denoting a MIT, even though it is present in the lattice solution at $n=1$. This is because the presence of a gap is not explicitly self-consistently determined, but rather the hole/particle spectral moments, which are indeed identical at convergence. A MIT is however found for $n=2-4$, given by the observed discontinuity in double occupancy at this point.

\begin{figure}[h!]
\includegraphics[scale=0.5,trim={0 0 0 0},clip]{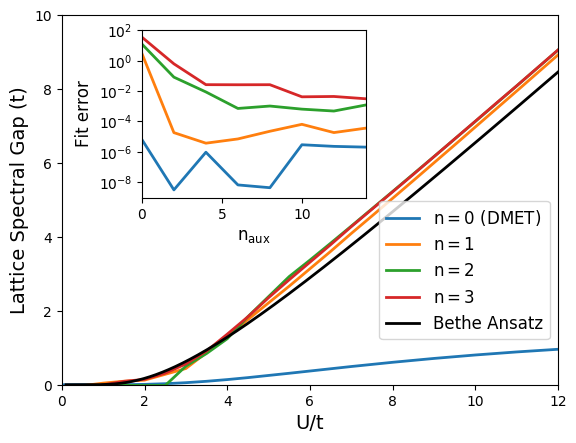}
\caption{Lattice spectral gap of the paramagnetic 1D Hubbard lattice with increasing hole/particle moment self-consistency and $n_{\rm imp}=2$, $n_{\rm aux}=10$.
Fitting $n\ge1$ retrieves a qualitative correct lattice gap in comparison to exact Bethe Ansatze results. Inset: Auxiliary hamiltonian fit errors for different $n$ as $n_{\rm aux}$ increases at $U=8t$.
}
\label{fig:Hubbard_Gap}
\end{figure}

\begin{figure}[h!]
\includegraphics[scale=0.5,trim={0 0 0 0},clip]{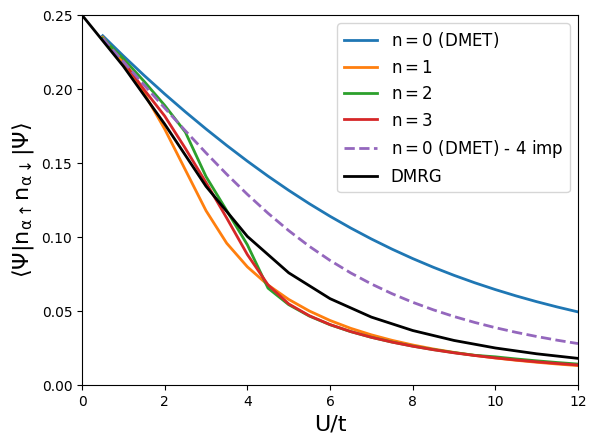}
\caption{Double occupancy of the impurity from the cluster solution for different self-consistent moment order $n$ for the paramagnetic 1D Hubbard model with $n_{\rm imp}=2$ (where no MIT should be observed), and including comparison to exact DMRG. $n_{\rm imp}=4$ DMET results are shown for comparison, demonstrating the improvement due to increased impurity size compared to improved dynamical resolution.}
\label{fig:Hubbard_DO}
\end{figure}

\section{One-dimensional Hubbard Chain}

The description of this opposing limit to the Bethe lattice is far from exact in cluster methods. This is even the case for complete dynamical self-consistency in DMFT (or $n, n_{\rm aux} \rightarrow \infty$ here) where for $n_{\rm imp}=1$ a spurious finite-$U$ phase transition is observed\cite{0953-8984-21-48-485602}. However, these results can be systematically improved with increasing $n_{\rm imp}$, while comparison to exact DMRG\cite{DMRG} (finite chains) or Bethe Ansatze\cite{Lieb68} (infinite system) results will enable us to benchmark the benefit of increasing dynamical self-consistency. 
We move to a two-site, half-filled paramagnetic impurity cluster, and consider the lattice spectral gap (Fig.~\ref{fig:Hubbard_Gap}) and cluster double occupancy (Fig.~\ref{fig:Hubbard_DO}). These are substantially in error for the DMET ($n=0$) description, while the inclusion of the dynamical effects can substantially improve upon the native DMET without the need for spin symmetry-breaking. The lattice spectral gap (Fig.~\ref{fig:Hubbard_Gap}) was estimated by considering the gap between eigenvalues about the chemical potential with more than 10\% of their weight on the physical (rather than auxiliary) sites. 


In contrast to the $n_{\rm imp}=1$ Bethe lattice results, larger numbers of auxiliary sites were required to converge the moment fit error (see Fig.~\ref{fig:Hubbard_Gap} inset), and results are shown for spectral gap and double occupancy at $n_{\rm aux}=10$ where the fit error is negligibly small.
Improving upon the DMET results, self-consistency with $n \ge 1$ is crucial to obtain a realistic gap when compared to Bethe-Ansatz results, where the spectral gap is dominated by the mean values of the particle and hole spectra defining the Hubbard bands. Improvements for higher moments are not significant for this property, with $n=2$ even showing some unphysical low-$U$ phase transition behaviour which disappears again for higher $n$\cite{0953-8984-21-48-485602}. The double occupancy of the cluster solution also shows modest systematic improvements as the dynamical effects are increased, with the large-$U$ behaviour particularly erroneous in the $n=0$ limit when compared to exact DMRG results. As the results converge with $n$, it is likely that further improvements to the exact limit are only going to be possible with increased impurity size and momentum resolution. An indication of this is given by the improvement in the $n=0$, $n_{\rm aux}=0$ (DMET) results, as we increase the impurity size to four sites, as shown in Fig.~\ref{fig:Hubbard_DO}. With this, the error in the double occupancy is approximately halved. However, despite this improvement, the error is still larger than including the low-order dynamical effects on the $n_{\rm imp}=2$ impurity size. This is true even at $n=1$, where the correlated cluster problem is half the size of the $n_{\rm imp}$=4 calculation, demonstrating the importance of including this partial dynamical information.

\section{Conclusion}

We have presented a computationally efficient quantum cluster method to systematically include effective finer dynamical resolution within a ground-state wave function embedding approach. This exactly and algebraically maps the full lattice with a given order of effective dynamical character to a finite cluster problem, whilst ensuring that the self-consistency can be rigorously satisfied back on the lattice via an optimized auxiliary space. In this way, we allow for any desired interpolation between the physics of DMET and DMFT, whilst remaining in an efficient wave function picture and rigorously finite impurity+bath cluster model. At the cost of doubling the dimensionality of this cluster problem to be solved, either the impurity space can be doubled, or two orders more lattice dynamical effects can be described. This allows for a well-posed, low-cost framework, where any desired trade-off in the resolution of momentum and dynamical effects of the correlation can be chosen for the problem at hand.

\section{Acknowledgements}

We would like to acknowledge fruitful discussions with Cedric Weber, Martin Eckstein, Garnet Chan, Stephen Clark and David Jacob in the course of this work. G.H.B. gratefully acknowledges support from the Royal Society via a University Research Fellowship, and has received funding from the European Union's Horizon 2020 research and innovation programme under grant agreement No. 759063.

\end{document}